# First-principles Study of Phonon Lifetime and Low Lattice Thermal Conductivity of Monolayer γ-GeSe: A Comparative Study

Bowen Wang[1], Xuefei Yan[1,2,3], Xiangyue Cui[1], Yongqing Cai[1, *]

[1]*Joint Key Laboratory of the Ministry of Education, Institute of Applied Physics and Materials Engineering, University of Macau, China*

[2]*School of Microelectronics Science and Technology, Sun Yat-sen University, Zhuhai 519082, China;*

[3]*Guangdong Provincial Key Laboratory of Optoelectronic Information Processing Chips and Systems, Sun Yat-sen University*

E-mail: yongqingcai@um.edu.mo

**ABSTRACT** Germanium selenide (GeSe) is a unique two-dimensional (2D) material showing various polymorphs stable at ambient condition. Recently, a new phase with a layered hexagonal lattice (γ-GeSe) was synthesized with ambient stability and extraordinary electronic conductivity even higher than graphite while its monolayer is semiconducting. In this work, via using first-principles derived force constants and Boltzmann transport theory we explore the lattice thermal conductivity ($\kappa_l$) of the monolayer γ-GeSe, together with a comparison with monolayer α-GeSe and β-GeSe. The $\kappa_l$ of γ-phase is relatively low (5.50 W/mK), comparable with those of α- and β- phases. The acoustic branches in α-GeSe are well separated from the optical branches, limiting scattering channels in the phase space, while for β-GeSe and γ-GeSe the acoustic branches are resonant with the low-frequency optical branches facilitating more phonon-phonon scattering. For γ-GeSe, the cumulative $\kappa_l$ is isotropic and phononic representative mean free path (rMFP) is the shortest (17.07 nm) amongst the three polymorphs, indicating that the $\kappa_l$ of the γ phase is less likely to be affected by the size of the sample, while for $\alpha$-GeSe the cumulative $\kappa_l$ grows slowly with mean free path and the rMFP is longer (up to 20.56 and 35.94 nm along zigzag and armchair direction, respectively), showing a stronger size-dependence of $\kappa_l$. Our work suggests that GeSe polymorphs with overall low thermal conductivity are promising contenders for thermoelectric and thermal management applications.

# I. INTRODUCTION

The remarkable thermoelectric, electrical, and optical properties [1-6] of two-dimensional (2D) materials have triggered extensive interest in recent years. Among the various 2D materials, group IV-VI compounds monochalcogenides (GeS, GeSe, SnS, SnSe, etc.) are a unique family with rich ambient stable phases and polar vibrations allowing potential optoelectronic applications [7-11]. The IV-VI monochalcogenides normally have a puckered layered structure similar to phosphorene and show versatile thermoelectric and ferroelectric transitions [12-15]. Meanwhile, due to the large structural space and $sp^3$ hybridization, the IV-VI chalcogenides have been experimentally demonstrated for fabricating phase change memory devices[16, 17].

As a member of the family of IV−VI monochalcogenides, germanium selenide (GeSe) normally adopts a typical puckered layered lattice (α-GeSe) and was fabricated for photodetector with a marked photoresponse to near infrared light illumination. Monolayer α-GeSe semiconductor was predicted with an anisotropic and low thermal conductivity[18, 19] and the phonon thermal transport of these puckered layered structures was widely studied[20, 21]. In addition to the puckered layered structure, a six-membered-ring structure of GeSe (β-GeSe) with uncommon boat conformation with $Pmn2_1$ symmetry (Fig. 1) was synthesized at high pressure and temperature and is stable under ambient conditions [22]. More recently, another phase of GeSe (γ-GeSe) with a four-atomic-thick hexagonal structure was predicted[23] and later synthesized[24]. The point group and space group of identified bulk γ-GeSe are $C_{6v}$ (6mm) and $P6_3mc$ respectively, while those are $C_{3v}$ (3m) and $P\bar{3}m1$ for monolayer γ-GeSe. Interestingly, γ-GeSe in bulk form possesses a small band gap but with an electronic conductivity even higher than graphite[23]. This high conductivity was found to be appealing for applications in lithium ion batteries, and Li atoms show a fast intercalation allowing for chemically exfoliation of γ-GeSe nanosheet[25]. Meanwhile, the electrical conductivity is several times higher than other 2D layered crystals which is highly promising for thermoelectric applications. For this novel hexagonal phase, many mysteries still remain unexplored. Besides the electronic properties, the thermal properties such as thermal conductivity of γ-GeSe, extremely important for the thermal management and application, are still unknown. Actually, except for monolayer α-GeSe, the phonon transport parameters of experimentally identified GeSe phases are still missing. Comparable studies of the various polymorphs of GeSe and the knowledge of the underpinning phonon transport features are the key to expanding the spectrum of their applications in optoelectronics and thermoelectric.

In this work, focusing on the new member of GeSe, we aim to investigate the phonon thermal properties of different experimentally identified phases of GeSe in the monolayer limit. Through deriving the Grüneisen parameter ($\zeta$) and phonon lifetime, anharmonicity of phonons and scattering due to phonon-phonon interactions are obtained. We find that for α-GeSe the acoustic phonons contribute about ~80% to the thermal conductivity while its contribution reduces for β-GeSe (~60%) and γ-GeSe (~70%). The magnitude of the thermal conductivity from large to small follows α-GeSe (zigzag) > γ-GeSe > β-GeSe (zigzag) > β-GeSe (armchair) > α-GeSe (armchair). For all the three GeSe phases, their thermal conductivities are overall lower than many other 2D materials, such as $MoS_2$, $MoSe_2$, silicene, $WS_2$, h-BN and black phosphorene. Finally, we provide some hints about the size-limited thermal conductivity which is critical for nanostructured materials and devices.

## II. COMPUTATIONAL METHOD

First-principles calculations based on the density functional theory (DFT) in conjunction with projector-augmented-wave (PAW) pseudopotentials [26] were performed using the Vienna ab initio simulation package (VASP) [27]. The exchange-correlation functional was treated using Perdew-Burke-Ernzerhof (PBE) type [28] of generalized gradient approximation (GGA). We used the same calculation settings as follows for all the three different phases of GeSe. The plane-wave energy cutoff was set as 400 eV. Along the out-of-plane direction, a necessary 20 Å thickness of vacuum space was used to avoid interactions between layers. A well-converged Monkhorst–Pack k-point grid 21 × 21 ×1 was taken to sample the Brillouin Zone. The convergence threshold was set as $10^{-8}$ eV for structural relaxation and electronic self-consistent calculations, and the cell volume and shape were fully optimized until the force tolerance is no larger than $10^{-4}$ eVÅ$^{-1}$.

Due to the semiconducting nature of the three phases in the monolayer [22, 23], phonons are the primary heat carriers and dominate the total thermal conductivity. Therefore, we mainly focus on the lattice thermal conductivity in this work. We have calculated the lattice thermal conductivity of different phases of monolayer GeSe by solving the linearized phonon Boltzmann transport equation (BTE) based on an adaptive smearing approach to the conservation of energy [29]. A full iterative solution[30] with ShengBTE code [31] was used.

The lattice thermal conductivity can be expressed as [31, 32]

$$\kappa_l^{\alpha\beta} = \frac{1}{k_B T^2 \Omega N} \sum_\lambda f_0(f_0+1)(\hbar\omega_\lambda)^2 v_\lambda^\alpha F_\lambda^\beta \qquad (1)$$

where $k_B$, $T$, $\Omega$, $N$ and $\hbar$ is the Boltzmann constant, temperature, volume of the unit cell, regular grid of q points and Planck constant respectively. $f_0$ is the distribution of phonons based on Bose-Einstein statistics and $\lambda$ is a phonon mode comprises branch index $p$ as well as a wave vector $\boldsymbol{q}$. $\alpha$ and $\beta$ are Cartesian index. $\omega_\lambda$ and $v_\lambda$ are the angular frequency and group velocity along $\alpha$ direction. Considering two- and three- phonon scattering as only scattering source, the linearized BTE $F_\lambda^\beta$ can be written as

$$F_\lambda^\beta = \tau_\lambda^0 (v_\lambda + \Delta_\lambda) \qquad (2)$$

Here $\tau_\lambda^0$ is phonon lifetime of mode $\lambda$ and $\Delta_\lambda$ is corrective term from iteration can be expressed as

$$\tau_\lambda^0 = N(\sum_{\lambda'\lambda''}^{+} \Gamma_{\lambda\lambda'\lambda''}^{+} + \sum_{\lambda'\lambda''}^{-} \frac{1}{2}\Gamma_{\lambda\lambda'\lambda''}^{-} + \sum_{\lambda'} \Gamma_{\lambda\lambda'})^{-1} \qquad (3)$$

$$\Delta_\lambda = \frac{1}{N}(\sum_{\lambda'\lambda''}^{+}\Gamma^{+}_{\lambda\lambda'\lambda''}(\xi_{\lambda\lambda''}\tau_{\lambda''}-\xi_{\lambda\lambda'}\tau_{\lambda'})+\sum_{\lambda'\lambda''}^{-}\frac{1}{2}\Gamma^{-}_{\lambda\lambda'\lambda''}(\xi_{\lambda\lambda''}\tau_{\lambda''}+\xi_{\lambda\lambda'}\tau_{\lambda'})) \quad (4)$$

where $\xi_{\lambda\lambda'} = \omega_{\lambda'}/\omega_\lambda$ and $N$ is the number of wave vector $q$ in the Brillouin zone. $\Gamma^{+}_{\lambda\lambda'\lambda''}$ and $\Gamma^{-}_{\lambda\lambda'\lambda''}$ are the absorption and emission of phonons[33-36]. $\Gamma_{\lambda\lambda'}$ is the scattering caused by impurity and boundary scattering[37, 38]. To get the harmonic and third-order interatomic force constants (IFC), 7 × 7 × 1 and 4 × 4 × 1 supercells were used respectively, with a 3 × 3 × 1 Monkhorst-Pack mesh of k points. The 14-th nearest neighbors were examined and included to get a convergent result. A 100 × 100 × 1 mesh was applied in the reciprocal spaces to get a convergent thermal conductivity. Details of convergence test are attached in Supporting Information (Fig. S1 to Fig. S5).

## III. RESULTS

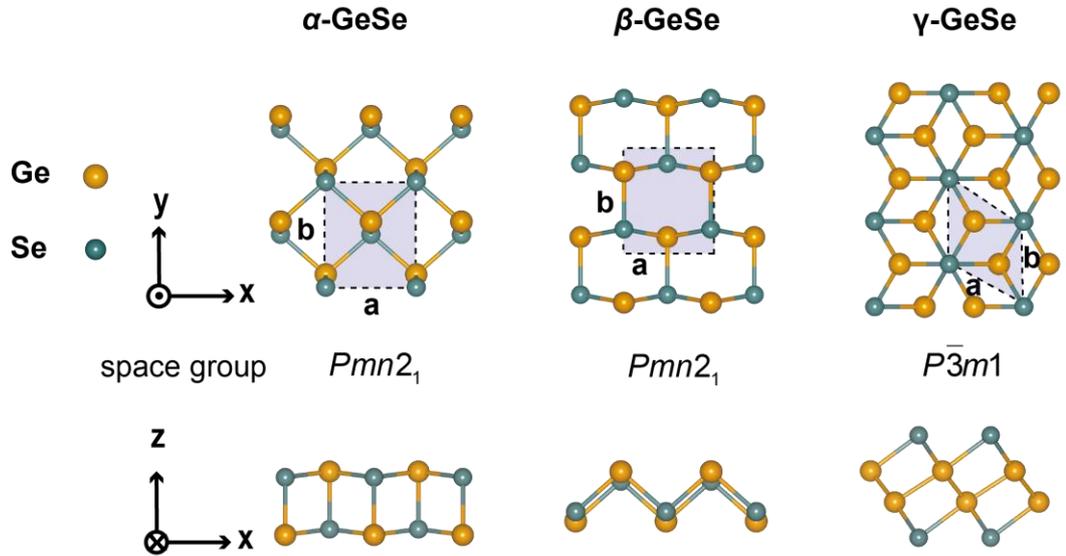

FIG. 1 Geometric structures of various phases of monolayer GeSe. From left to right, α-GeSe, β-GeSe, and γ-GeSe are plotted.

We firstly compare the atomic structures of the three GeSe polymorphs. The structure of α-GeSe with a puckered conformation is composed of buckled six-rings with vertically alternating Ge and Se atoms. *β*-GeSe has a boat conformation for its six-rings as shown in Fig. 1. Distinct from the *α*-GeSe and *β*-GeSe, *γ*-GeSe has a hexagonal structure with four-atomic-thick Se-Ge-Ge-Se sublattice and can be considered as two merged buckling honeycomb lattices. The optimized lattice parameters are listed in Table 1. which consistent with the experimentally and theoretically reported values[19, 22, 39]. The calculated phonon dispersion curves are shown in Fig. 2 which fits well with the result reported [19, 22, 23] and no imaginary frequencies exists, indicating the monolayer structures of three phases GeSe are dynamically stable. Due to the four atoms in primitive cell, there are 12 phonon modes (three acoustic modes and nine optical modes) in monolayer GeSe. The longitudinal acoustic (LA) and transverse acoustic (TA) branches are linear and the flexural acoustic (ZA) branch is quadratic in proximity to the $\Gamma$ point which commonly exists in 2D materials[40-44].

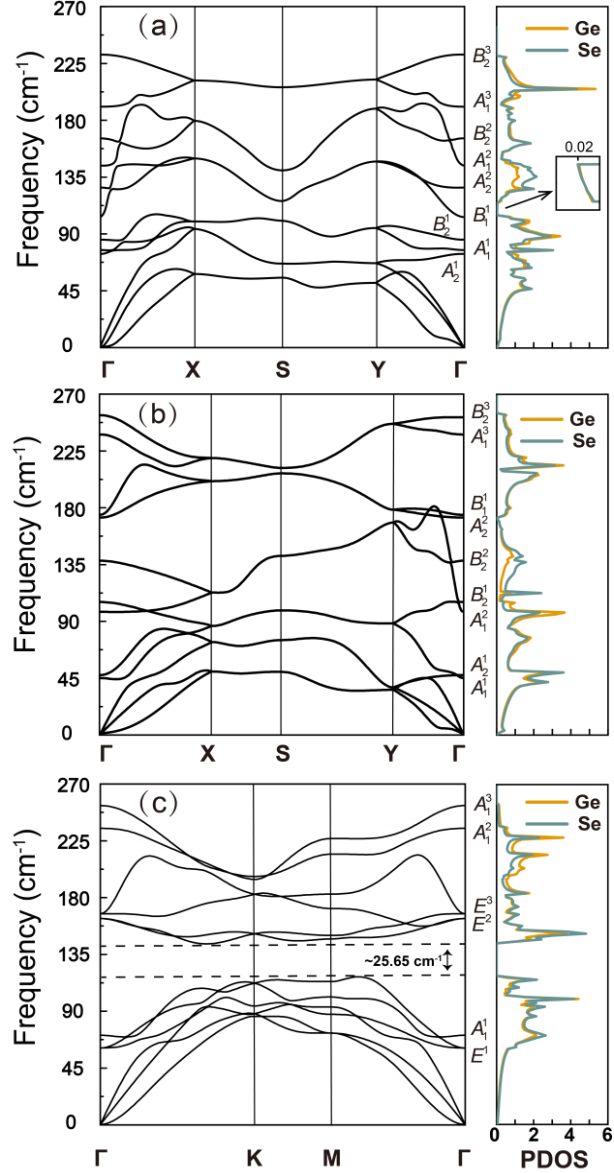

FIG. 2 Phonon dispersion curves and phonon density of states (PDOS) of monolayer (a) *α*-GeSe, (b) *β*-GeSe, and (c) *γ*-GeSe.

In addition to the similarity, we also note that some differences are found amongst the three polymorphs. Firstly, for the *γ*-GeSe, the low-frequency optical branches (< ~ 120 cm$^{-1}$) cross over the acoustic branches (Fig. 2). In addition, there exists an appreciable phononic gap with width of 25.65 cm$^{-1}$ formed between acoustic branches and mid-frequency optical branches (~140 cm$^{-1}$). Interestingly, there is no similar gap in *α*-GeSe and *β*-GeSe. A relatively large bandgap between acoustic (a) and optical (o) modes induces limited phonon scattering via (a, a, o) process and thus large phonon lifetimes. In contrast, a tiny or absent bandgap leads to a relatively high phonon scattering rate and a small phonon lifetime. Second, the optical branches of *β*-GeSe are notably dispersive, thus inducing considerable group velocities of optical phonon branches. Such broad distribution of phononic branches of *β*-GeSe, together with a gapless phonon dispersion, renders more scattering path and possibility in phase space than its cousins. Third, the point group of *γ*-GeSe is hexagonal $C_{3v}$, which is different from the orthogonal $C_{2v}$ of *α* and *β*-GeSe. The calculated eigenvectors and frequencies of the optical modes at Γ are depicted in Fig. 3. According to group theory, for *γ*-GeSe the optical phonon modes at Γ point are decomposed as:

$$\Gamma_{optical}(\gamma-GeSe) = E^{1,2,3}(IR+R) + A_1^{1,2,3}(IR+R) \qquad (5)$$

where $E^{1,2,3}$ and $A_1^{1,2,3}$ are doubly degenerate horizontal and singly degenerate vertical vibration modes respectively. IR and R represent the infrared and Raman active nature, respectively. From Eq. (5), we note that all the optical modes in monolayer *γ*-GeSe are IR and R active which is quite different from stacked *γ*-GeSe[24].

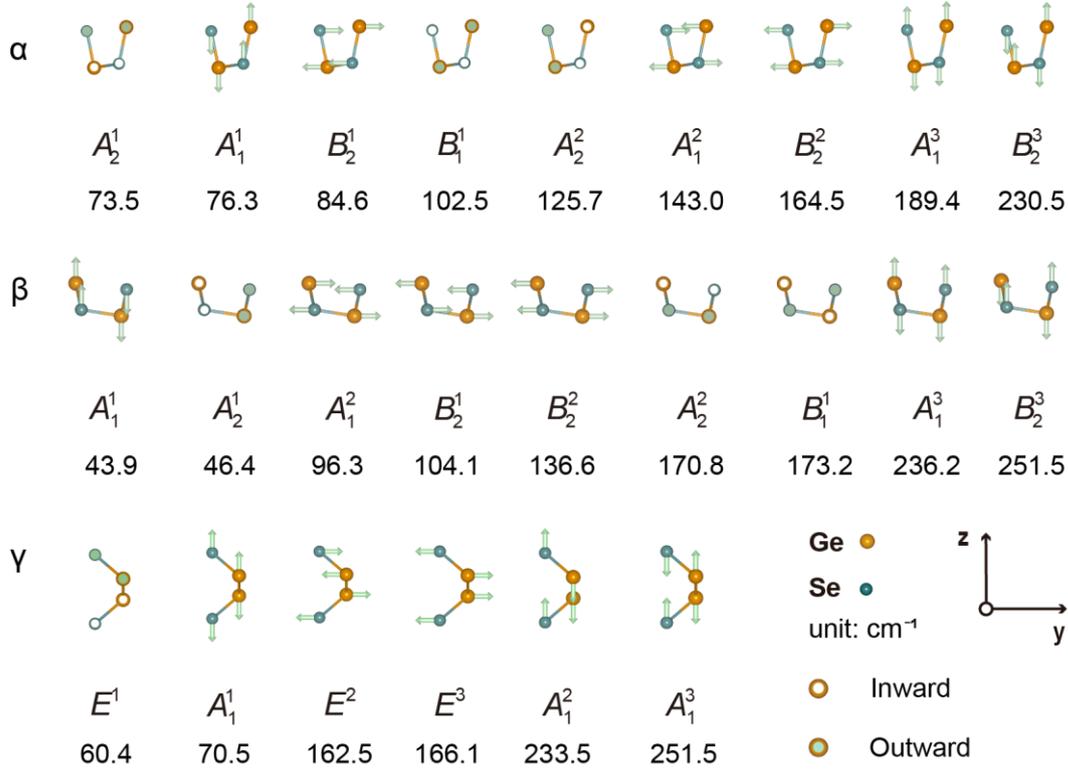

FIG. 3. The optical phonon modes and its vibration eigenvectors of monolayer α, β and γ GeSe at Γ of Brillouin zone.

The calculated lattice thermal conductivity $\kappa_l$ of α, β and γ-GeSe monolayers as a function of temperature is shown in Fig. 4. Due to the Umklapp phonon scatterings, the differences in the values of intrinsic $\kappa_l$ of different phases GeSe diminish as temperature rises. For α-GeSe, as expected and consistent with previous work [19] and other hinge-like structure materials [45], its thermal transport is anisotropic and the $\kappa_l$ at room temperature (300 K) is 5.94 W/mK (4.12 W/mK) along armchair (zigzag) direction. The group velocities of LA branch ($v_{LA}$) along zigzag (Γ to X) and armchair (Γ to Y) direction are obtained by analyzing the slope of LA branches near the Γ point. The orientational dependence of $\kappa_l$ correlates with anisotropic group velocities $v_{LA}$ of 3.38 (2.56) km/s for zigzag (armchair) direction. For β-GeSe, surprisingly, with showing an apparent anisotropic structure (Fig. 1b), the anisotropy of $\kappa_l$ is not obvious and the $\kappa_l$ along zigzag direction (4.29 W/mK) is slightly higher than armchair direction (4.16 W/mK). In the case of γ-GeSe, due to the isotropic hexagonal structure, it exhibits intrinsically isotropic thermal properties with the $\kappa_l$ of 5.50 W/mK. The exact values of the $\kappa_l$ and $v_{LA}$ for the three phases are compiled in Table 1. For comparison, the order of $\kappa_l$ from large to small follows α-GeSe (zigzag) > γ-GeSe > β-GeSe (zigzag) > β-GeSe (armchair) > α-GeSe (armchair) as listed in Table 1. The values of $\kappa_l$ are overall lower than many other 2D materials, such as $MoS_2$, $MoSe_2$, silicene, $WS_2$, h-BN and black phosphorene[46-49]. Recent studies[50, 51] have revealed that even at room temperature, the higher-order phonon scattering process, such as four-phonon scattering, is crucial to the thermal transport in broad bandgap semiconductors. It is worth noting that a significant phononic band gap (25.65 $cm^{-1}$) separates the 12 phonon branches into two parts, associated with the four atoms in the γ-GeSe primitive cell, with the upper half consisting of 6 optical branches and the lower part of 3 acoustic branches and 3 optical branches. The gap will limit the processes of two acoustic phonon combining to one mid-frequency optical phonon, accordingly reducing the interaction of optical and acoustic phonons. The four-phonon processes between the optical and acoustic phonons would be unlikely to be constrained by this band gap, which may result in a lower $\kappa_l$ of γ-GeSe than the current value with considering only two and three phonon scatterings.

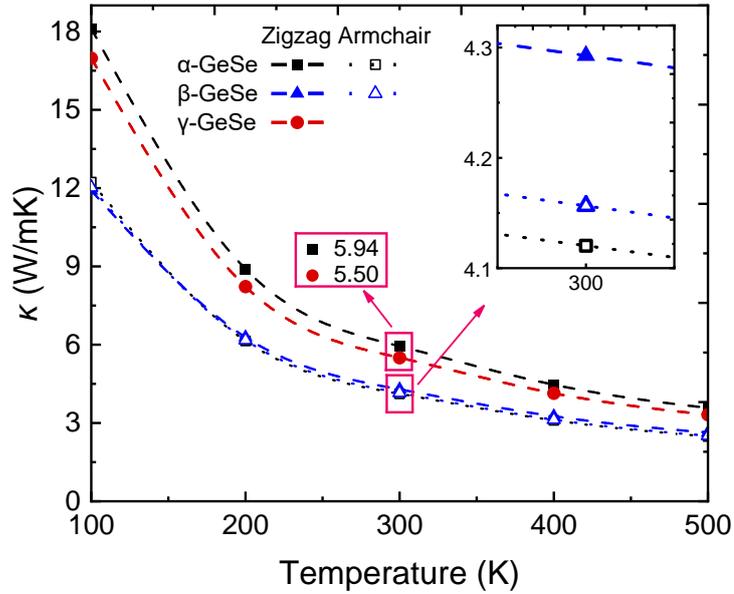

FIG. 4 Thermal conductivity ($\kappa$) of monolayer (α, β and γ) GeSe as a function of temperature along zigzag (solid symbol) and armchair (open symbol) directions.

Table 1. Lattice constants, $\kappa_l$ at 300 K, phonon group velocities of LA branches ($v_{LA}$) at Γ point, contribution of phonon branches to $\kappa_l$, and rMFP at 300 K of monolayer α-GeSe, β-GeSe and γ-GeSe.

| Phase | Direction | Lattice constant (Å) | $\kappa_l$ (W/mK) | $v_{LA}$ (km/s) | Phonon branches contribution (%) | | | | rMFP (nm) |
|---|---|---|---|---|---|---|---|---|---|
| | | | | | ZA | TA | LA | Optical | |
| α-GeSe | zigzag | 3.98 | 5.94 | 3.38 | 30.03 | 22.37 | 27.54 | 20.06 | 20.56 |
| | armchair | 4.27 | 4.12 | 2.56 | 34.53 | 25.74 | 20.67 | 19.06 | 35.94 |
| β-GeSe | zigzag | 3.67 | 4.29 | 3.19 | 28.26 | 17.64 | 15.10 | 39.0 | 10.76 |
| | armchair | 5.90 | 4.16 | 2.38 | 41.10 | 15.88 | 10.81 | 32.21 | 18.82 |
| γ-GeSe | Isotropic | 3.81 | 5.50 | 4.43 | 37.84 | 18.95 | 15.91 | 27.3 | 17.07 |

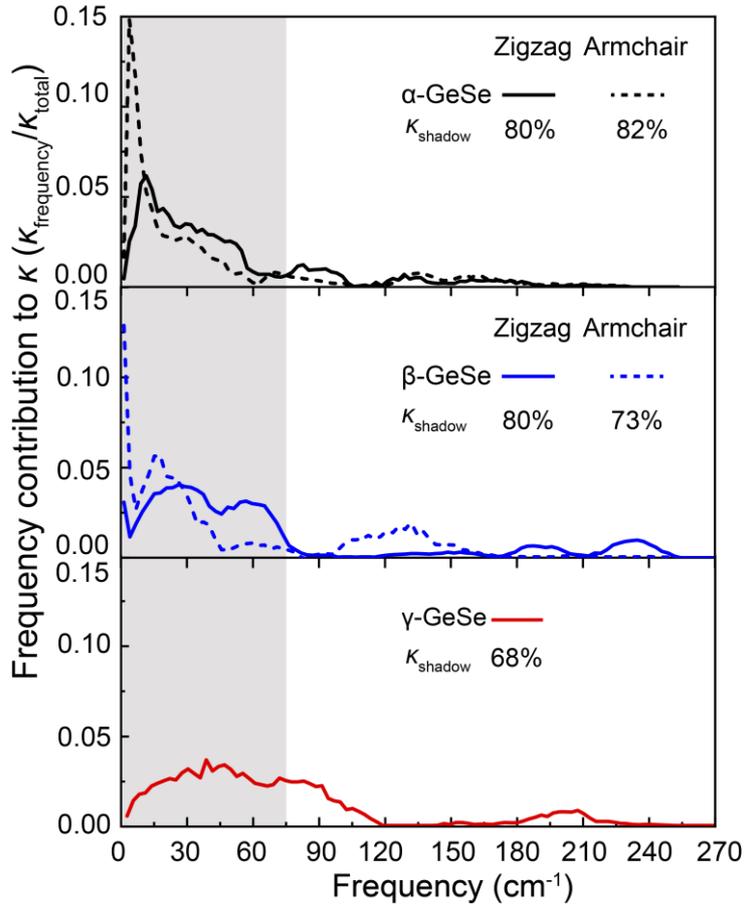

FIG. 5 The phononic contribution to $\kappa_l$ of monolayer (α, β and γ) GeSe as a function of frequency of modes. $\kappa_{\text{shadow}}$ corresponds to the weight of contribution of modes with frequencies in the shadowed window.

As shown in Fig. 5 and statistics, a large portion (~80%) of the $\kappa_l$ is contributed by the low-frequency phonons ranging from 0 to 75 cm$^{-1}$ in $\alpha$ and $\beta$ GeSe. This range is the typical characteristic acoustic phonons (Fig. 2) with spectrally different phonon lifetime and group velocity as well as can be seen below. Due to the crossover and hybridization of optical and acoustic branches, some dispersions of phonons ranging from 0 to 90 cm$^{-1}$ are relatively flat. The contributions of acoustic and optical phonons to the thermal conductivity at 300 K are analyzed and listed in Table 1.

For the $\alpha$-GeSe, the acoustic phonons make the dominant contribution (~80%) to the $\kappa_l$. This could be ascribed to a weaker scattering of acoustic modes and thus leading to a significant amount of heat conduction higher than $\beta$-GeSe and $\gamma$-GeSe. In the case of $\alpha$-GeSe, the acoustic bands are well separated from the optical branches (Fig. 2a), thus reducing the scattering channel in the phase space. In contrast, for $\beta$-GeSe and $\gamma$-GeSe, the acoustic branches are resonant with the low-frequency optical branches, leading to promoted scatterings and a reduced contribution of the acoustic modes to the $\kappa_l$. For all the cases, we found that the ZA mode is more efficient to convey the heat current than the LA and TA modes.

Notably, for $\beta$-GeSe, the optical modes make a remarkable contribution up to 39% (32%) to the $\kappa_l$ along the zigzag (armchair) direction (Table 1), much higher than $\alpha$- and $\gamma$-GeSe. This could be due to its most dispersive nature of phononic branches, especially for optical modes greater than 180 cm$^{-1}$, as indicative of the doublet peaks in Fig. 5. As shown in Fig. 6a, $\beta$-GeSe shows a strongly higher phononic group velocity for the intermediate optical modes between 100 and 200 cm$^{-1}$ than the other phases.

To understand the mechanism of the relatively low $\kappa_l$ of different phases of monolayer GeSe, we calculated $\zeta$ and phonon lifetime as shown in Fig. 6b and c, respectively. The mode-resolved Grüneisen parameter $\zeta(qv)$ at the wave vector $q$ and the band index $v$ can be expressed as

$$\zeta(q v) = -\frac{V}{2[\omega(q v)]^2} \langle e(q v)|\frac{\partial D(q)}{\partial V}|e(q v)\rangle \qquad (6)$$

where $V$, $\omega(qv)$, $D(q)$ and $e(qv)$ is the cell volume, phonon frequency, dynamical matrix, and the eigenvector respectively. The phase space for three-phonon scattering (P3) with an inverse connection with $\kappa_l$ [36] is also investigated. As we can see from the Fig. S6, the order of P3 at low frequencies is almost indistinguishable among the three phases, resulting the little difference in the contribution of low-frequency modes to $\kappa_l$ among the phases (shown in Table 1). For the high-frequency part, the P3 of β- and γ-GeSe is obviously lower than α-GeSe, leading to a low scattering rate at high frequency and a higher contribution to $\kappa_l$ from optical modes.

According to Fig. 6b, all the three phases of GeSe exhibit a negative $\zeta$ and small group velocities at low-frequency range, leading a strong anharmonicity and a low $\kappa_l$. The order of phonon lifetime at low frequency follows monolayer α-GeSe > β-GeSe > γ-GeSe (Fig. 6c). For β-GeSe, the area with significant negative $\zeta$ spans almost the full zone of the low-frequency modes (ω < 45 cm$^{-1}$) which is different from α-GeSe and γ-GeSe, suggesting the strongest anharmonicity among the three phases. This is consistent with our previous analysis that there exist gapless and highly dispersive phonon branches, leading to more scattering path in phase space in the β-GeSe. The difference of the symmetry (space group Pmn2$_1$ for α/β-GeSe and P$\bar{3}$m1 for γ-GeSe) and the bonding type may mainly be responsible for the distinct magnitudes of $\zeta$ between α/β-GeSe and γ-GeSe. Although the group velocities of β-GeSe of the optical phonon branches in the mid-frequency area (100-200 cm$^{-1}$) are the largest across a broad spectrum due to the notably dispersed optical branches in phonon dispersion (Fig. 6a), the phonon lifetime of β-GeSe (around 1 ps) is relatively short owing to the strong phonon-phonon scattering. The strongly shortened lifetime for both the high-frequency optical phonons (> 30 cm$^{-1}$) and low-frequency acoustic branches is responsible for the lowest $\kappa_l$ in β-GeSe.

For γ-GeSe, its phonon lifetime spans from 0.1 to 10 ps. In particular, the low-frequency acoustic phonon (< 30 cm$^{-1}$) has the shortest lifetime compared with its two other cousins (Fig. 6c). Our calculation indicates that these low-frequency phonons, associated with long-wavelength acoustic vibrations of atoms, are more likely to be scattered compared with $\alpha$- and $\beta$-GeSe. The underlying reason could be related to its unique thick-sandwiched structure with Se-Ge-Ge-Se where the mismatch of mass and atomic radius between the atoms in the planar inner core Ge layer and outer Se layers hinders the flow of heat via these low-energy phonons. This is similar to the strong scattering of low-energy rigid interlayer phonons experienced in the bilayer or multilayers materials[52, 53]. Another reason can be related to its strongly overlapping between the acoustic branch and the low-energy optical branch (60-120 cm$^{-1}$) as shown in Fig. 2c. This dramatically increases the multi-phonon scattering assisted by low-energy acoustic phonons and accordingly a low relaxation time.

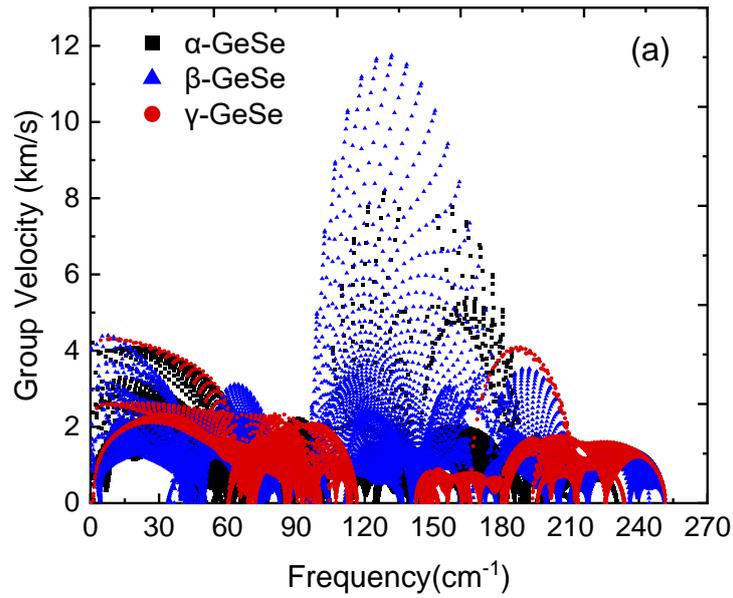

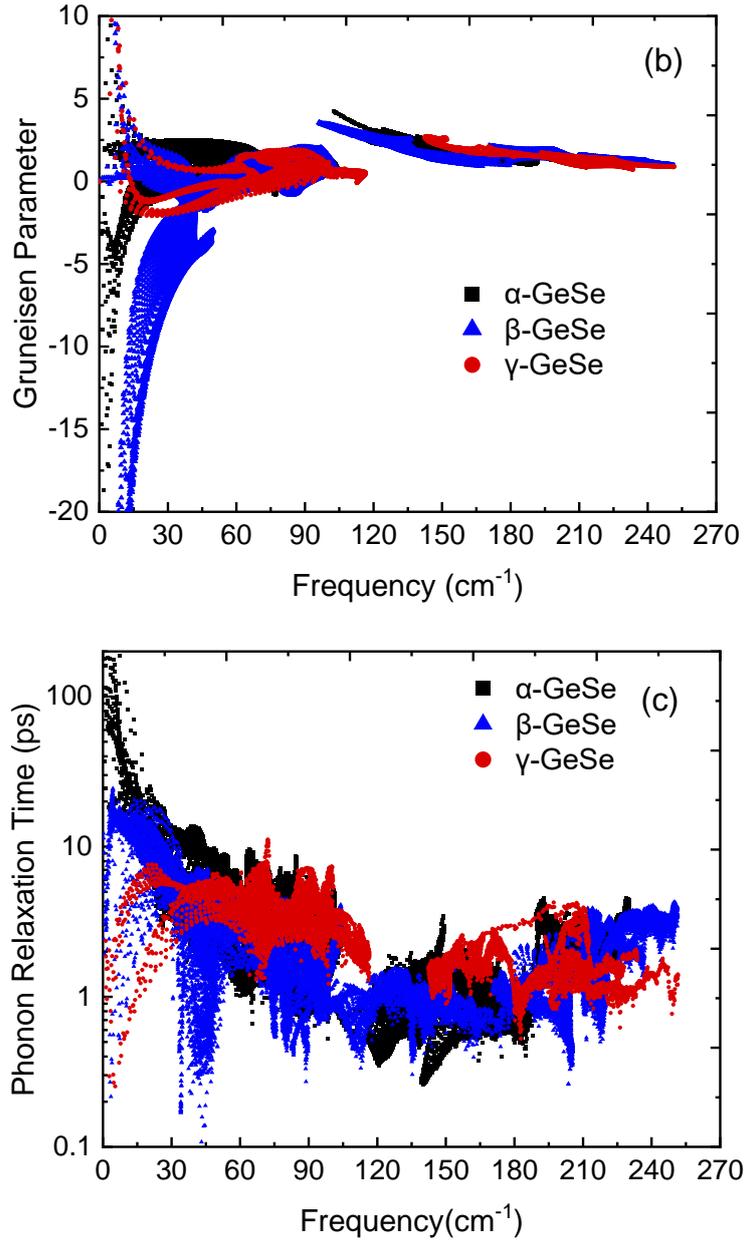

FIG. 6 (a) phonon group velocity, (b) Grüneisen parameter (ζ), (c) phonon lifetime of monolayer *α*-GeSe, *β*-GeSe and *γ*-GeSe as a function of frequency at 300 K.

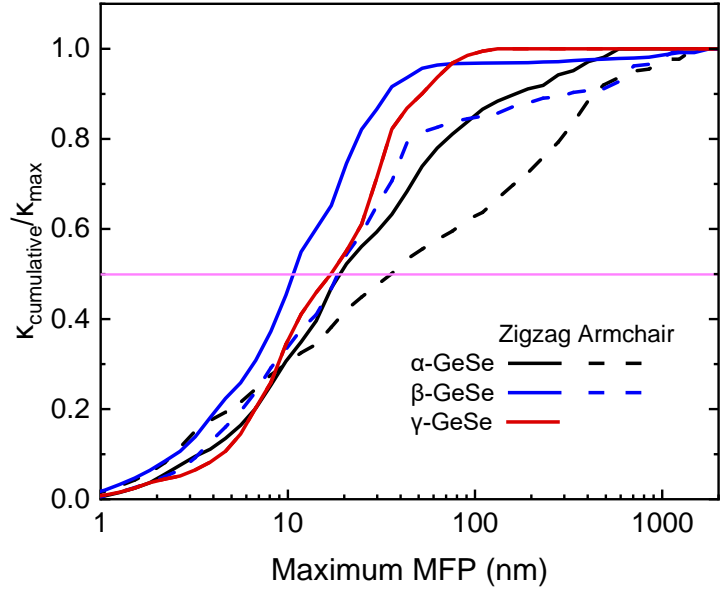

FIG. 7 Cumulative thermal conductivity along zigzag and armchair directions of α, β and γ-GeSe as a function of mean free path (MFP) at 300 K.

We next provide some insights about size dependent behaviors of $\kappa_l$ for the three polymorphs. The cumulative $\kappa_l$ with a mean free path below a threshold value $l_{max}$ can be fitted as:

$$\kappa_l(l \leq l_{max}) = \frac{\kappa_0}{1 + l_0 / l_{max}}$$

where $\kappa_0$ is final cumulated $\kappa$ and $l_0$ is the parameter assessed by fitting. The cumulative $\kappa_l$ as a function of phonon mean free path (MFP) is calculated and shown in Fig. 7. These results show a specific value of $\kappa_l$ (normalized to respective maximum value) contributed by phonons with MFP up to a certain value, acting as an approximation of the size-limited $\kappa_l$ of nanostructured GeSe. They can be measured and compared with the thermal-conductivity spectroscopy technique [54, 55]. The cumulative $\kappa_l$ rises with the MFP, then gradually saturates. Therefore, by effectively reducing the MFP such as via intentional doping, the effective $\kappa_l$ of the system can be reduced. Experiments studies[56, 57] showed that the thermal conductivity of doped GeSe is significantly reduced due to the presence of impurities.

The representative mean free path (rMFP), defined as the MFP value corresponding to half of the convergent $\kappa_l$, is useful metrics allowing a quantitative evaluation about the size effect on phonon thermal transport. The calculated values of rMFP of three phases GeSe are listed in Table 1, which are quite small compared to MoS$_2$[58], resulting in lower $\kappa_l$ of the GeSe phases. It is noted that the $\kappa_l$ of the β-GeSe along different directions is nearly the same. However, the rMFP of β-GeSe is ~10.76 and 18.82 nm in the zigzag and armchair directions, respectively which means a prominent anisotropy of the $\kappa_l$ of nanosized β-GeSe. As we can see from Fig. S6, when the MFP is less than 100 nm, the anisotropy (κ$_{zigzag}$/κ$_{armchair}$) of monolayer β-GeSe is relatively high, but it drops to roughly 1 as the MFP increases. For the monolayer α-GeSe, when the MFP grows about 100 nm, the anisotropy increases to the highest and subsequently declines until it approaches a constant value (~1.5).

## IV. CONCLUSIONS

In summary, we have systematically investigated the phonon properties and lattice thermal conductivity of monolayer α-, β- and γ-GeSe using the first-principles method and phonon Boltzmann transport equation. To understand the mechanism of the low thermal conductivity, we calculated the phonon group velocity, Grüneisen parameter, phonon lifetime, and phase space volume. The three polymorphs possess very different phonon dispersions, leading to a subtle difference in phonon scattering and conduction. Amongst the phases, α-GeSe has the well-separated and continually distributed acoustic and optical modes giving rise to the longest lifetime and the highest thermal conductivity. β-GeSe has the most dispersive phonons which is accompanied with the highest group velocity for optical modes. The optical phonons make a remarkable contribution to the thermal conduction up to 40%. Compared to these two phases, the polymorph with the strongest anharmonicity exists in β-GeSe, leading to the lowest thermal conductivity. γ-GeSe has the shortest lifetime of acoustic modes owing to a strong overlapping with low-energy optical modes and a unique sub-layered Ge and Se lattice. The thermal conductivity of γ-GeSe is isotropic and about 5.50 W/mK. This paper gives a thorough investigation of phonon transport in different phases (*α*, *β* and *γ*) monolayer GeSe, which will be important for GeSe-based thermal management and thermoelectric application.

**Supporting Information:** Convergence test; Phase space volume of α-GeSe, β-GeSe and γ-GeSe as a function of frequency; Anisotropy of thermal conductivity ($\kappa_{zigzag}/\kappa_{armchair}$) of α-GeSe, β-GeSe and γ-GeSe as a function of MFP.

**ACKNOWLEDGEMENTS**

This work was supported by the National Natural Science Foundation of China (Grant 22022309), and Natural Science Foundation of Guangdong Province, China (2021A1515010024), University of Macau (SRG2019-00179-IAPME, MYRG2020-00075-IAPME), and Science and Technology Development Fund from Macau SAR (FDCT-0163/2019/A3). This work was performed at the High Performance Computing Cluster (HPCC), which is supported by the Information and Communication Technology Office (ICTO) of the University of Macau.